# A-Site Cation disorder and Size variance effects on the physical properties of multiferroic $Bi_{0.9}RE_{0.1}FeO_3$ Ceramics (RE = $Gd^{3+}$, $Tb^{3+}$, $Dy^{3+}$)

T. Karthik [†,§], T. Durga Rao [†], A. Srinivas [‡], and Saket Asthana [†*]

[†]*Advanced Functional Materials Laboratory, Department of Physics, Indian Institute of Technology Hyderabad, Andhra Pradesh – 502205, India.*

[§]*Department of Materials Science and Engineering, Indian Institute of Technology Hyderabad, Andhra Pradesh – 502205, India.*

[‡]*Advanced Magnetics Group, Defence Metallurgical Research Laboratory, Kanchanbagh, Andhra Pradesh – 500058, India.*

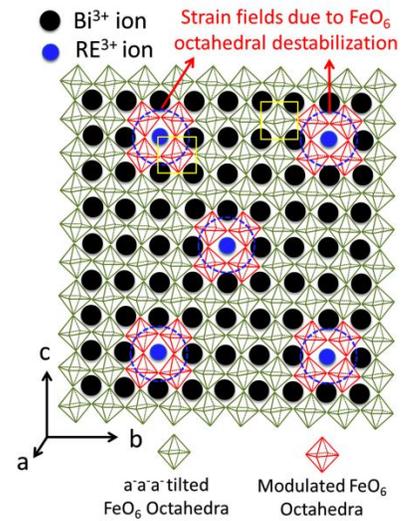

**ABSTRACT:** Lattice disorder effects due to A-site size variance between $Bi^{3+}$ and $RE^{3+}$ (RE= Gd, Tb, Dy) cations in polycrystalline $BiFeO_3$ (BFO) ceramics substituted with an invariable $RE^{3+}$-ion concentration of 10mol% evidenced a coexistence of orthorhombic Bragg reflections along with the *R3c* phase. Reduced tolerance factor values for RE-BFO samples indicate the coexistence of both in-phase and anti-phase tilting system. Raman scattering experiments elucidated the change in covalency of Bi-O bond and a selective softening of phonon mode corresponding to oxygen motion suggests a significant destabilization in the $FeO_6$ octahedra. FE-SEM micrograph depicts a disparity in the microstructure with reduced grain size in RE-BFO samples. EDAX analysis confirmed the stoichiometry of all samples. Optical responses studied in a spectral range from 1.5 eV to 6 eV were dominated by three Charge transfer transitions ($Fe_1$ 3d→ $Fe_2$ 3d, $O_p$ → $Fe_d$, $O_p$ → $Bi_p$) and a doubly degenerate d-d transition ($^6A_{1g}$ → $^4T_{2g}$) in all the samples. $RE^{3+}$-ion substitution in BFO host lattice red shifts the position of d-d transition band, which indicates the variation in crystal field strength. The suppression of spiral spin structure release the macroscopic magnetization so called weak ferromagnetism in RE-BFO samples, was evidenced from isothermal Magnetization measurements. Heat flow measurement reveals a decrease in Neel transition temperature ($T_N$) with $RE^{3+}$-ion substitution endorses the suppression of antiferromagnetic ordering. Present results also provide an evidence for the weakening of long range ferroelectricity with $RE^{3+}$-ion substitution, which was confirmed from the suppression of endotherm, assigned for Curie transition temperature ($T_C$) around 820°C. As a consequence of large A-site size variance, the change in phase transition temperatures for RE-BFO samples were explained on the basis of strain fields established by ordered oxygen displacements. A decrease in vacant sites was confirmed from the lower density of states $N(E_f)$ of charge carriers in RE-BFO samples, further it also correlates well with the hopping length ($R_{min}$) and conductivity studies.

## INTRODUCTION

Among the multiferroic oxide, $BiFeO_3$ (BFO) is probably the most extensively studied system, due to its intrinsic coexistence of the two ferroic order parameters (i.e., magnetic and ferroelectric) well above the room temperature. Due to these interesting phenomena BFO has received enormous attention in the scientific community to explore their functional properties for possible electronic/technological applications.[1,2]

At room temperature, bulk $BiFeO_3$ (BFO) crystallizes in a rhombohedrally distorted perovskite system with *R3c* space group.[3] In this structure due to the $6s^2$ lone pair mechanism, both $Bi^{3+}$ and $Fe^{3+}$cations were displaced along the three fold $[111]_P$ polar axis and off-centered with respect to barycenter of the oxygen polyhedra, which inturn gives rise to ferroelectricity.[3,4] Further, the rotation of adjacent oxygen $FeO_6$ octahedra around the [111]$_P$ direction gives rise to the antiphase tilting system (a$^-$a$^-$a$^-$) corresponding to Glazers notation.[5] As a consequence, BFO has a very high ferroelectric ordering Curie transition temperature (T$_C$) around ~830°C with a large ferroelectric remnant polarization (P$_r$) being around 50-150 μC/cm$^2$ for high quality BFO thin films[6-8] and 2.5-40 μC/cm$^2$ for Bulk BFO ceramics[9-12], along the polar <111> direction. In the case of magnetic ordering, Fe$^{3+}$ magnetic moments due to partially filled 3d-orbitals has a G-type canted antiferromagnetic ordering with an incommensurate space modulated spiral spin structure along the [110]$_P$ direction with a long periodic wavelength of 620Å.[13] As a result of space modulated spiral spin structure, a residual non-zero magnetization was observed with an antiferromagnetic ordering Neel transition temperature T$_N$ around ~372°C.[14] Despite these interesting behaviors it has some drawbacks such as leakage current, formation of secondary phases, cation vacancies due to Bi-volatilization, anion vacancies due to oxygen etc., has hampered its practical applications. In order to reduce these drawbacks chemical substitution on their respective cationic sites (Bi/Fe site) has been studied widely by several researchers globally. Among them rare earth (RE) ion substitution at A-site got extensive attraction, because it gives rise to the evolution of weak ferromagnetism (WFM)[12, 15, 16] and improved electrical properties.[17, 18] A-site substitution in BFO with RE$^{3+}$-ion is known to introduce structural distortions owing to its smaller ionic radii and it gives us a wide range of platform to explore their structure-property correlations. Interestingly, a general transition from R3c symmetry to orthorhombic Pnma space group of the prototype GdFeO$_3$ structure occurs for RE-ion substituted BFO ceramic. However, based on the RE-ion size and concentrations of the substituent, several intermediate symmetries between the two end members have been reported.[12, 19, 20] Some recent reports illustrate that Gd substitution in BFO induce a polar to polar R3c →Pn2$_1$a structural phase transition for x=0.1 and for higher concentrations polar to non-polar Pn2$_1$a→Pnma phase transition occurs.[21] But other authors reported a coexistence of R3c and Pbnm phases for a similar composition.[22] Further, Neutron diffraction experiments on Tb substituted BFO also confirms the existence of bi-phasic model (R3c + Pnma) beyond a concentration, x = 0.1.[23] Similarly for Nd and Sm substitutions in BFO also exhibits a coexistence of Pnma phase beyond x ≥0.1.[24] However in most of the reports a coexistence of two phases (i.e., a bi-phasic model) has been observed when the concentration of RE ion exceeds beyond ~7.5 -10%. In most of the reports, authors have focused on the characterization of physical properties; but the effects of RE substitution for Bi$^{3+}$ ion in the host lattice and their structure-property relationship have been little studied. Also the change in phase transition temperatures has a lot of controversies, such as decrease /increase in T$_N$ and T$_C$ values for different compositions.[23, 24, 25] Further, no strong explanations were evidenced for these behaviors. Here, in this present work we have studied the role of coexisting orthorhombic phases on the physical properties by choosing an invariable RE$^{3+}$-ion (Gd, Tb, Dy) concentration of 10%. We have explained the change in phase transitions (i.e., T$_N$ and T$_C$) for RE-substituted BFO sample based on strain fields created by ordered oxygen displacements using Attfield's model. Also, structure-property correlations with the coexisting orthorhombic phases along with BFO (R3c space group) were established.

[*]Corresponding Author e-mail: asthanas@iith.ac.in, Tel: +91-40-2301 6067.

## EXPERIMENTAL SECTION

Polycrystalline samples of BFO and RE-ion substituted BFO ceramics (Bi$_{0.9}$RE$_{0.1}$FeO$_3$; RE = Gd, Tb, Dy) were prepared by conventional mixed oxide solid state reaction technique using high purity precursor oxides Bi$_2$O$_3$, Fe$_2$O$_3$, Gd$_2$O$_3$, Tb$_2$O$_3$, Dy$_2$O$_3$ (99.99% Sigma Aldrich Chemicals) respectively. All the precursors were pre-dried for 10 hrs at 400°C to check the loss of ignition. After pre-drying the precursors were weighed according to desired cation ratios and milled for 3 hrs using IPA as a medium. These milled powders were calcined at 700°C for 1 h and further grinded finely, sieved and then compacted into rectangular pellets followed by final sintering at 800°C for 3 hrs in air atmosphere with a heating rate of 5°C/min. The phase analysis of sintered samples was examined by powder x-ray diffractometer (PANlytical x'pert pro) using a cu kα incident radiation (λ=1.5406 Å) with a 2θ step size of 0.016° over the angular range 20° ≤ 2θ ≤ 80°. Raman scattering spectra were measured at room temperature using a Laser Micro Raman spectrometer (Bruker, Senterra) in a back-scattering geometry with an excitation source of 785nm. The power of laser spot on the polished sample surface was 10mW, so that sample heating was insignificant. The data were collected with each increment of 0.5cm$^{-1}$ and an integration time of 10s. Diffuse reflectance spectra (DRS) of the powders were recorded in the wavelength range from 200-900nm using (Shimadzu UV-3600) UV-Vis-NIR spectrophotometer having a wavelength accuracy of ± 1nm resolution. Microstructural analysis was carried out on the fractured surface of pellets by FE-SEM (Carl Zeiss-ULTRA 55). EDAX Measurements were performed using HITACHI, S-3400 SEM. Room temperature magnetization (M vs H) measurements were performed using PPMS with VSM assembly (Quantum Design, USA) upto a field of 5 Tesla. The sintered discs were Polished then coated with silver paste and annealed at 400°C for 30 min. Room temperature Dielectric behavior of the silver electroded samples were determined using a precision impedance analyzer (Wayne Kerr-6500B, U.K). The ferroic order transitions (i.e., magnetic and ferroelectric) were determined through Heat flow measurements using Differential scanning calorimetry (DSC). The Neel transitions were measured using a TA instruments DSC-Q200 having a temperature accuracy of +/- 0.1°C at a controlled heating rate of 10°C/min under N$_2$ atmosphere. Ferroelectric transition temperature and other high temperature transitions of each composition were measured using NETZSCH-STA-449 F3 high temperature DSC. Before performing DSC measurements, the instrument was calibrated with standard indium samples and the measurements were repeated twice to ensure the consistency of results.

## RESULTS AND DISCUSSION

Figure 1 shows the powder X-ray diffraction patterns of BiFeO$_3$ (BFO) and heavy rare earth ion substituted BFO ceramics such as Bi$_{0.9}$Gd$_{0.1}$FeO$_3$ (BFO-Gd), Bi$_{0.9}$Tb$_{0.1}$FeO$_3$ (BFO-Tb), and $Bi_{0.9}Dy_{0.1}FeO_3$ (BFO-Dy). It was observed that BFO stabilizes in *R3c* symmetry which matches well with the JCPDS card no: 71-2494 and along with BFO some low intense secondary phases such as $Bi_2Fe_4O_9$ and $Bi_{25}FeO_{40}$ were also observed. The formation of such secondary phases is unavoidable during the kinetics of BFO formation. Similarly, it was found that RE-BFO ceramics also stabilizes in rhombohedral system with *R3c* symmetry. But on close observation on the major Bragg peaks as shown in Figures 2(a) to (d) there found some extra reflections superimposed along with the *R3c* phase. These reflections cannot belong to the secondary phases (i.e., $Bi_2Fe_4O_9$/$Bi_{25}FeO_{40}$), because there were no such reflections observed at this 2θ position in the un-substituted BFO ceramic. This confirms that these extra reflections in RE-BFO ceramics arise as a result of the superposition of two spectral phases possibly due to the ortho-rhombic phase contribution. Several recent reports also showed orthorhombic phase coexistence with $Pn2_1a$ /*Pnma* /*Pbnm* space group along with BFO (*R3c*).[19, 21-24] This formation depends solely on the ionic radii of RE-ion and their concentration of substitution.

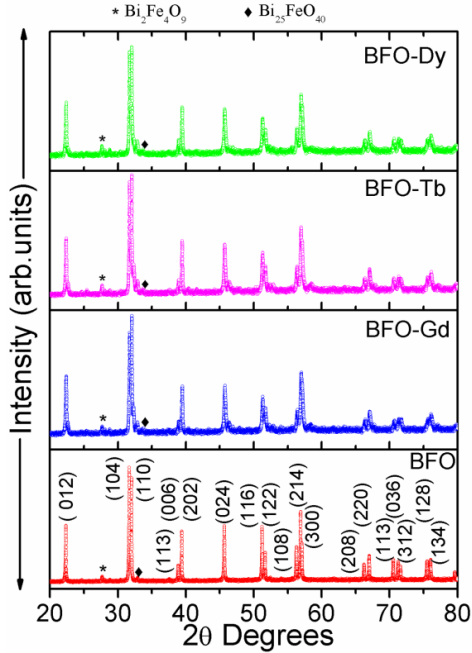

Figure 1. X-ray diffraction patterns of sintered BFO, BFO-Gd, BFO-Tb and BFO-Dy ceramics.

The origin of these orthorhombic phases in RE-BFO samples can be explained on the basis of tolerance factor which is associated with the cationic size variance between $Bi^{3+}$ and RE-ions. Goldschmidt proposed a factor (t) to quantify size mismatch between A and B cations to the cubic perovskite ($ABO_3$) topology.[26] Tolerance factor (t) is defined as, $t = (\langle r_A \rangle + r_O)/\sqrt{2}(r_B + r_O)$, Where, $\langle r_A \rangle$ is the average ionic radii of A-cation in 12-fold coordination, $r_B$-ionic radii of B-cation in 6-fold coordination, $r_O$-ionic radii of Oxygen anion in 6-fold coordination respectively.[26, 27] When t is in unity, it corresponds to an ideal perovskite with undistorted structure, while t<1 has a distorted perovskite system indicating tilt or rotation of the $BO_6$ octahedra. Based on extensive structural data of perovskites, Reaney et al.[29] showed that at room temperature perovskites with 0.985 < t < 1.06 are expected to have untilted structures. Perovskites with 0.964 < t < 0.985 usually have anti-phase tilted structures and perovskites with t < 0.964 are expected to show both in-phase and anti-phase tilting[28, 29]. As t continues to decrease, the stability of the perovskite system decreases and eventually it will not form. Several authors reported the tolerance factor value by considering Shannon ionic radii of $Bi^{3+}$ ion (r = 1.17 Å) in 8-fold coordination[30], if that is the case t will be 0.8886. Hence according to Reaney et al. it will not stabilize in the perovskite structure.[29]

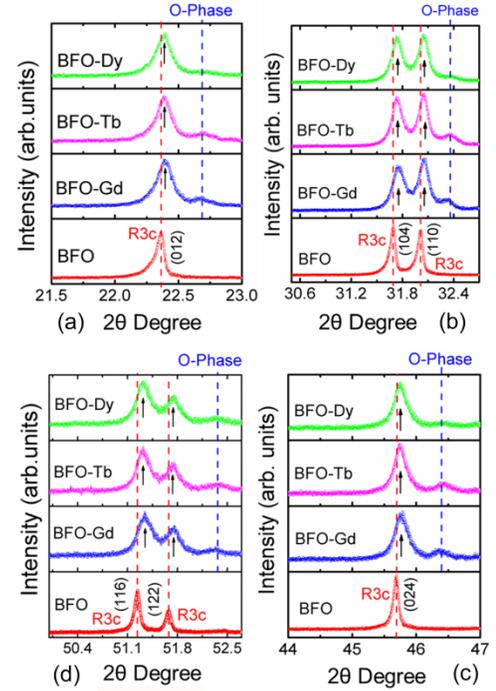

Figure 2(a) to (d) indicates the close observation of Bragg reflections corresponding to the major planes of BFO and RE-BFO samples. Arrow (→) indicates the corresponding shift in 2θ position for RE-BFO samples. Dotted red line (···) indicates the Bragg reflections corresponding to R3c phase and blue dotted line (···) indicates the appearance of orthorhombic phase in RE-BFO samples.

However, several authors reported a t value of 0.96 for BFO [31, 32, 24]. In this context by considering 12-fold co-ordination (C.N) ionic radii for $Bi^{3+}$ ion (r = 1.40Å), then a t value of 0.9682 was observed for BFO. This value matches well within the limit of anti-phase tilted perovskite system as proposed by Reaney et.al. Hence it is worthwhile to consider the ionic radii of 12-fold C.N for A-Cations (i.e., for $Bi^{3+}$, $RE^{3+}$) in BFO and RE-BFO ceramic system[33]. RE-ion substitution in BFO lattice declines the tolerance factor value from 0.9682 (BFO) to 0.9626 (BFO-Dy). Therefore, it clarifies that the substitution of $Gd^{3+}$, $Tb^{3+}$ and $Dy^{3+}$-ions in BFO introduces a change in the octahedral tilt system with reduced unit cell volume, because the ionic size of $Gd^{3+}$ = 1.27 Å (C.N: 12); $Tb^{3+}$ = 1.25 Å (C.N: 12), C; $Dy^{3+}$ = 1.24 Å (C.N: 12); is smaller than that of $Bi^{3+}$ ion[32]. As BFO has a t value of 0.9682 it exists in $a^-a^-a^-$ (Glazers notation) anti-phase titling system and as t decreases to 0.9637 for BFO-Gd system, it exhibits both anti-phase and in-phase titling system. As a consequence this reflects in the formation of orthorhombic phase within the global *R3c* symmetry. Thus the A-site cation size variance and tolerance factor (t) was also a responsible factor for the formation of orthorhombic phases along with *R3c* phase. Similarly on close observation of the Bragg reflections corresponding to major planes, a shift in their two theta positions (2θ) and increase in full width half maximum (FWHM) were observed remarkably. The right shift in 2θ position indicates a reduction in d-spacing which directly relates to the reduction in unit-cell volume. Hence we can strongly attribute that $RE^{3+}$-ion substitution in BFO lattice shrinks the unit cell volume. Further, a significant increase in FWHM as seen in Figures 2(a) to (d) of RE-BFO sample may possibly be due to the decrease in grain size, which is consistent with the SEM micrographs.

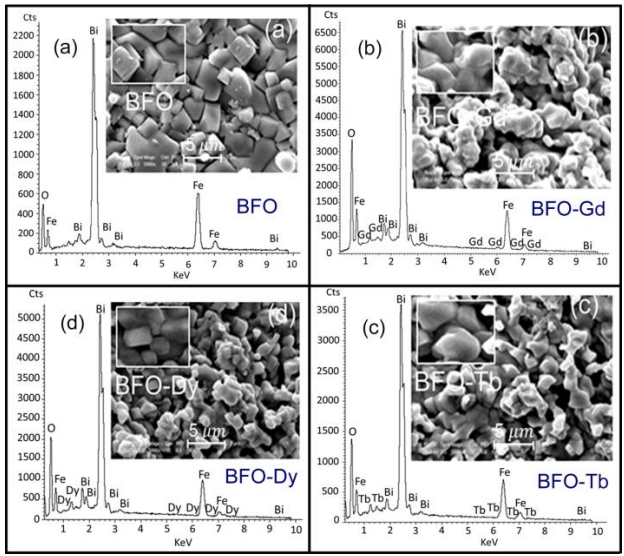

Figure 3. FE-SEM micrographs and EDAX analysis performed on the fracture sections of sintered (a) BFO, (b) BFO-Gd, (c) BFO-Tb and (d) BFO-Dy ceramics.

FE-SEM micrograph shown in Figures 3(a) to (d) of RE-BFO samples depicts the microstructural and morphological change as compared to the un-substituted BFO ceramics. In BFO ceramic the grains are found to be densely packed and regular in shape with an average grain size in the range between 3-5 µm. But in the case of RE substituted BFO ceramic, the morphology of grains is irregular and less dense with reduced grain size in the range between 1-2 µm. The less dense packing of grains in RE-BFO samples might be due to the lower sintering temperature (i.e., 800°C). Generally for RE-substituted BFO ceramic, several authors have increased the sintering temperature based on the concentration of substitution.[21, 22, 24] In order to perceive the exact change in properties, we have adopted identical sintering temperature and processing conditions for all the samples. EDAX analysis confirms the stoichiometry of all samples and found that RE-ion concentration in RE-BFO samples were about ~10mol%.

In our materials studied here, a slight structural disorder due to compositional variation was revealed from XRD investigations in RE-BFO samples. However, local structural distortions such as titling of octahedra and ionic displacements that are coherent over a short length scale about few unit cells could be easily missed in XRD investigations. Recent studies prove that Raman scattering could be an ideal technique which can probe local distortions of the crystal structure and ionic configurations on length scales substantially shorter than 10 nm.[34-37]

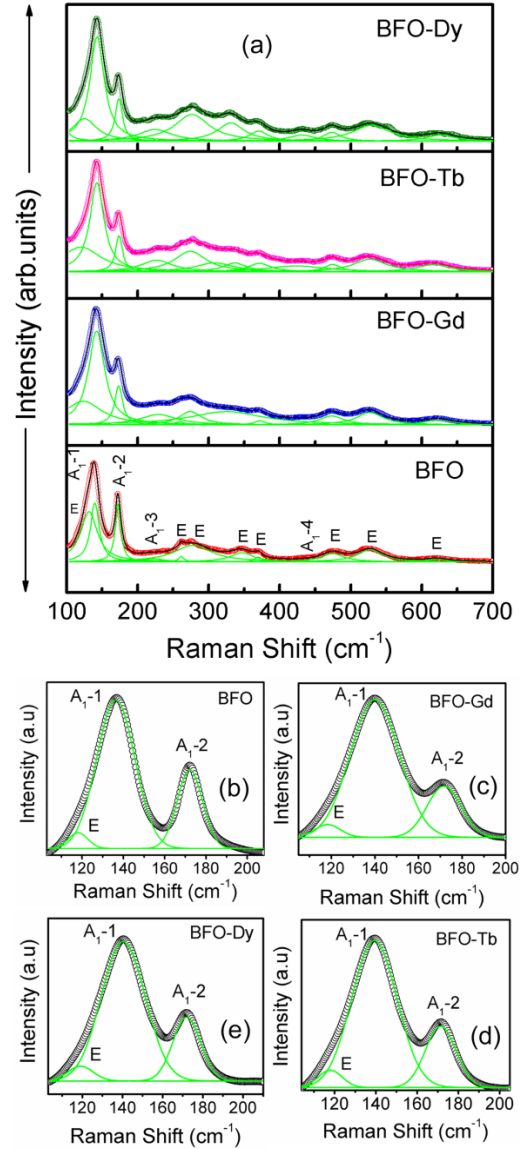

Figure 4(a) Raman scattering spectra of sintered BFO and RE-BFO samples (color open circles) measured at room temperature, together with their fitted spectra (— black thin solid line) and the de-convoluted Raman active modes (— solid green line). Figure 4(b) - (e) Spectral deconvolution for selected phonon modes of BFO and RE-BFO samples.

Figure 4 (a) shows the measured room temperature Raman scattering spectra of BFO and RE-BFO samples. By fitting the measured spectra and decomposing the fitted curves into individual Lorentzian components, the peak position of each component i.e., the phonon modes (cm$^{-1}$) which are Raman active was obtained for each samples. Rhombohedral $R3c$ ($C_{3v}$) structure of BFO has 10 atoms in the unit cell which gives rise to 27 optical phonon modes in the zone centre ($K \approx 0$). Among them 13 phonon modes (4A$_1$ + 9E) with A$_1$ and E symmetry are both Raman and IR active [38, 39]. The deconvoluted spectra of BFO in Figure 4(a) shows 12 active Raman modes including A$_1$-1 and A$_1$-2 modes with strong scattering intensities at 136.3 cm$^{-1}$ and 172.1 cm$^{-1}$, with a weak scattering intensity around 226.5 cm$^{-1}$ and 426 cm$^{-1}$ corresponding to A$_1$-3 and A$_1$-4 modes and 8 E modes with medium scattering intensities at 118, 261, 277.9, 346, 370.2, 474.7, 527.5 and 621.4 cm$^{-1}$ respectively were observed.

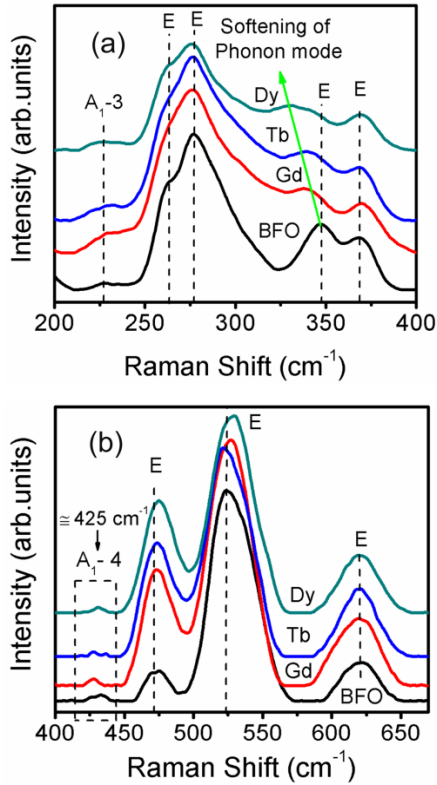

Figure. 5 (a) & (b) Close observation of the Raman active modes indicating weak scattering intensities of A$_1$-3, A$_1$-4 and arrow (→) indicating softening of phonon modes with RE-ion substitution in BFO lattice.

According to first principle calculation, Hermet et.al [39] suggests that Bi-atoms highly participate in low frequency modes below 167 cm$^{-1}$, while Fe atoms were involved in modes between 152 and 262 cm$^{-1}$ with possible contribution to higher frequency modes and the motion of oxygen atoms dominated by the modes above 262 cm$^{-1}$. The A$_1$ modes observed for BFO in our present work were in good agreement with that of M.K. Singh et.al[40], which manifests longitudinal-optical A$_1$ (LO) mode of the perovskite system. However, hardening of E modes was observed in our sample, which could be due to the possible oxygen vacancies persisting in the sample[40, 41]. In most of the reports, authors witnessed a strong scattering intensity of A$_1$-3 phonon mode but in our case a weak scattering intensity was observed. The reason for this weak scattering intensity of A$_1$-3 mode could be due to the higher excitation wavelength i.e., 785 nm (1.59eV). A similar kind of observation was reported by Y. Yang et.al[42] when they excited BiFeO$_3$ with different wavelengths.

Since Raman scattering spectra are sensitive to atomic displacements, the evolution of Raman modes for RE-ion substitution in BFO lattice can provide valuable information. As expected the spectral features of RE-BFO samples shows the 12 phonon modes which are the characteristic features of $R3c$ symmetry. On close observation of A$_1$-1 and A$_1$-2 modes as shown in Figure 4 (b) to (e), a drastic change in intensity of A$_1$-2 mode, broadening (FWHM) and shift in the phonon frequency of A$_1$-1 mode was noticed for RE-BFO samples. We can use a simple harmonic model to discuss the shift of A$_1$-1 phonon mode toward higher frequencies: $\omega = \sqrt{k/m^*}$, where $k$ is the force constant and $m^*$ the reduced mass [41, 43]. The valence and ionic radii of RE-ions (Gd$^{3+}$, Tb$^{3+}$, Dy$^{3+}$) are closer to that of Bi$^{3+}$ ion, hence there will not be much impact due to the force constant. However the lower atomic weight (i.e., reduced mass, $m^*$) of Gd (157.25 g/mol), Tb (158.92 g/mol) and Dy (162.5 g/mol) relative to Bi (208.98 g/mol) could be attributed to the increase in A$_1$-1 phonon mode frequency.

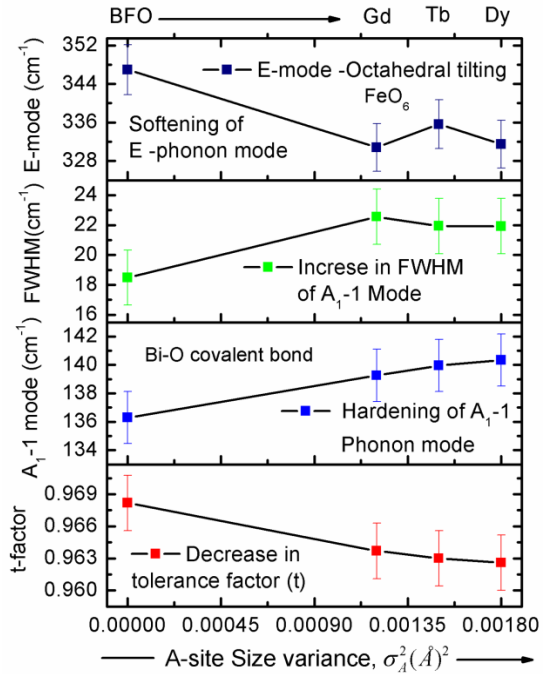

Figure 6. Variation of tolerance factor (t), frequency shift in A$_1$-1 Phonon mode, FWHM of A$_1$-1 Phonon mode and frequency shift in E-mode corresponding to oxygen motion with respect to A-site size variance plotted with standard deviation of each data as their error bar.

As a whole, a significant change in all these phonon characteristics (i.e., shift in phonon mode, intensity and FWHM) corresponding to A$_1$-1 and A$_1$-2 phonon mode clearly indicates that A-site disorder was induced due to RE-ion substitution at Bi$^{3+}$ site in the host BFO lattice. Similarly, on close observation of different phonon modes as shown in Figure 5 (a) & (b) depicts a weak scattering intensity of $A_1$-3 and $A_1$-4 mode. Further an abrupt softening about 10 cm$^{-1}$ in magnitude of E mode near $\cong$350cm$^{-1}$ associated to oxygen motion was observed in RE-BFO samples as depicted by arrow in Figure 5(a). This significant softening (i.e., decrease in frequency) in E-mode due to RE-ion substitution strongly perceives that there is a significant destabilization in the octahedral oxygen chains, which in-turn affects the local FeO$_6$ octahedral environment by titling of the oxygen octahedra coupled via octahedral rotation.[33, 44] Recent high pressure investigations on BFO also depicted a similar softening of E-modes attributed to the modulation in FeO$_6$ local environment due to high pressure.[44, 45] Figure 6 shows that RE-ion substitution in BFO induces A-site cation disorder due to the increase in size variance ($\sigma_A^2$) i.e., $\sigma_A^2 = \langle r_A^2 \rangle - \langle r_A \rangle^2$, which in-turn reduces the tolerance factor (t) values.[46, 47] As evidenced from Figure 6, an increase in A-site size variance ($\sigma_A^2$) and decrease in tolerance factor value affects directly the $A_1$-1 phonon mode corresponding to Bi–O covalent bond and E-modes corresponding to oxygen motion. From our Raman scattering results, it can be concluded that a significant A-site disorder accompanied by a change in octahedral tilt system could lead to the orthorhombic phase formation along with the *R3c* phase as such observed in X-ray diffraction results. Similar to a change in other phonon modes, a softening of E mode (i.e., 261.6 cm$^{-1}$) around ~3-5 cm$^{-1}$ in magnitude for RE-BFO sample was observed. This phonon mode was associated closely with Fe–O covalent bond; this indirectly relates to the possible change in Fe–O–Fe bond angle and further it correlates well with the origin of weak ferromagnetism in our RE-BFO samples.

Figure 7 (a) – (d) shows the room temperature UV-Visible absorption spectra of BFO and RE-BFO samples derived from the diffuse reflectance (R) spectrum using Kubelka-Munk function,[48] F(R) = (1-R)$^2$/2R which is plotted as a function of energy (eV). BFO has a distorted cubic structure (rhombohedral) hence there is a point group symmetry breaking from O$_h$ to C$_{3v}$ as shown in Figure. 7(e). In particular, by considering C$_{3v}$ local symmetry of Fe$^{3+}$ ions (3d$^5$ – High spin configuration; $t_{2g}^3 e_g^2$) in BFO and using the correlation group and subgroup analysis for the symmetry breaking from O$_h$ to C$_{3v}$, there expected to have six d-d transitions between 0 and 3eV.[49] From the DRS spectrum of BFO as shown in Figure.7(a) a shoulder centered around 1.92 eV corresponds to $^6A_{1g} \rightarrow {}^4T_{2g}$ excitation which arises due to the d-d crystal field excitations of Fe$^{3+}$ ions in BFO.[49, 50] Formally these excitations are forbidden because they change the total spin of Fe$^{3+}$ from S = 5/2 to S = 3/2 as depicted in Figure 7(f). However, spin orbit coupling relaxes the spin selection rule and gives rise to these transitions.[50, 52] Above 2.2 eV, absorption increases substantially and displays a small shoulder centered at 2.47eV and other two large features near 3.28 eV and 4.9 eV were assigned to the charge transfer (CT) excitations.[49] Among these CT transitions the band centered at 2.47 eV corresponds to two center CT transitions which is driven by Me$_1$ 3d – Me$_2$ 3d (i.e., Fe$_1$ 3d – Fe$_2$ 3d) intersite electron transfer. The other two bands around 3.28 eV and 4.9 eV corresponds to one center CT transitions which are associated with the interatomic p-d (O 2p - Fe 3d) or p-p (O 2p - Bi 5p) transitions.[51] A similar kind of absorption bands as such in BFO were also observed for the RE substituted BFO samples indicating their electronic energy level scheme looks similar. But on close observation, there appeared to be a slight redshift on the d-d and CT transition bands for RE-BFO samples. A noticeable shift of around 0.7-0.8 eV was observed in the d-d (i.e., $^6A_{1g} \rightarrow {}^4T_{2g}$) crystal field transition of RE-BFO samples as shown in Figure 7(e). The position of this d-d transition depends strongly on the Fe-O bond length, site symmetry and Fe-O-Fe exchange interaction[45]. A redshift in this transition energy is proportional to the increase of crystal field splitting Δ.[45] Susana G. S et al.[45] observed a similar redshift in $^6A_{1g} \rightarrow {}^4T_{2g}$ transition for BFO by a pressure induced spectroscopic study. This clearly indicates that RE-ion substitution in BFO increases the internal chemical pressure which result in the changes of FeO$_6$ local environment, as a consequence of contraction in unit cell volume and coexistence of orthorhombic phase due to the octahedral symmetry breaking from C$_{3v}$ to D$_{2h}$.

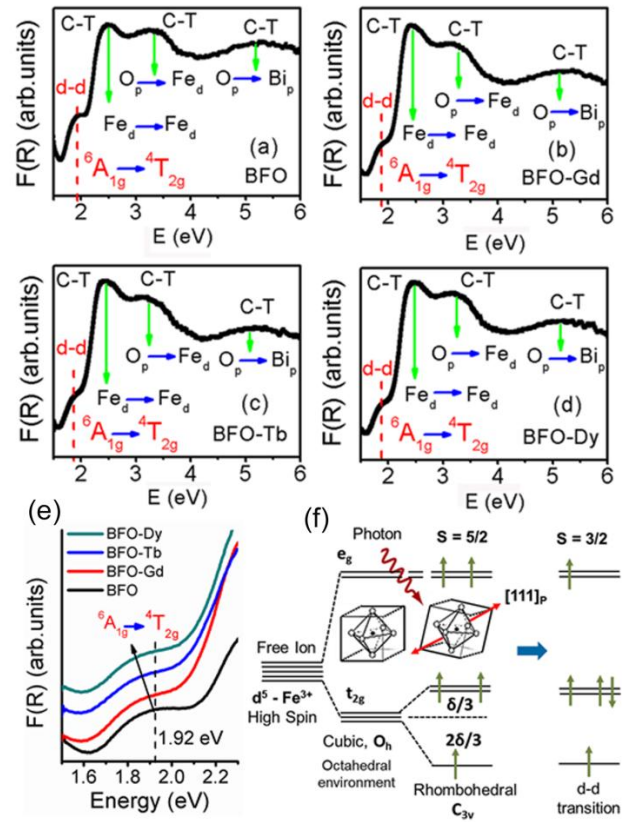

Figure 7(a) - (d) Room temperature Diffuse Reflectance UV-VIS spectroscopy plots depicting the electronic transitions of sintered BFO and RE-BFO samples. Figure 7(e) Redshift in d-d transition band indicated by arrow (→). Figure Symmetry breaking in BFO from O$_h$ to C$_{3v}$ and the spin orbit coupling stabilization gives rise to crystal field spin forbidden d-d transitions.

Room temperature magnetization (M-H) curves of BFO and RE-BFO samples measured up to a magnetic field of 5 Tesla were displayed in Figure 8 (a). BFO, is a G-type antiferromagnet[13] with a cycloidal spin arrangement incommensurate with its lattice over a period of 620Å. Hence it has no macroscopic magnetization at room temperature. Investigations of room temperature magnetic properties in RE-ion substituted BFO ceramics confirm the existence of weak ferromagnetism (WFM) due to the presence of non-zero remnant magnetization ($M_r$) and high coercivity ($H_C$) values as shown in Figure 8(b) and (c). In general, the origin of this WFM in RE-BFO sample is explained macroscopically due to suppression of space modulated spiral spin structure (SMSS).[19-22] Further, the intrinsic phenomenon for suppression of SMSS and evolution of WFM can be explained on the basis of following mechanisms:

(i) The structural distortions observed in RE-BFO sample can induce changes in the local environment such as Fe–O bond lengths and Fe–O–Fe bond angles. As evidenced from our Raman spectroscopy results, there observed some significant changes in the phonon mode near 350cm$^{-1}$ attributed to the modulation of $FeO_6$ octahedral tilts.[44] This strongly perceives that there are some possible changes in the Fe-O local environment. In general, when the tolerance factor value (t) is less than unity (i.e., t<1), the Perovskite ($ABO_3$) system adjusts by co-operative rotations of the $BO_3$ arrays and these rotations create $(180-\phi)$ B-O-B bond angles.[53] Further, the bending angle $\phi$ increases with a decrease in t value.[53] Since the t value for BFO is 0.9682, the $FeO_3$ array rotates along the $[111]_P$ axis and significantly bends the Fe-O-Fe bond angles close to around $\cong 155°$.[54, 55] In the case of RE-BFO sample, the bending angle $\phi$ increases further due to the reduced t values as discussed earlier. As a result, the antiferromagnetic super-exchange interaction ($J_{ij}$) between the Fe spins will decrease according to the given expression $H = -\sum_{i \neq j} J_{ij} S_i S_j$, where H is the Hamiltonian function and $J_{ij}$ is the exchange integral, $S_i$, $S_j$ are the spins at i and j-site.[56] Among the RE-BFO samples, Dy substituted BFO shows a significant WFM behavior which corresponds directly to the lower t value as compared to all the other samples. It can be inferred that the increase in bending angle $\phi$ leads to the evolution of WFM in RE-BFO samples at the expense of SMSS suppression.

(ii) Other than the change in Fe-O-Fe bond angle, the relevant possible mechanism for WFM can be explained on the basis of the evolution of Dzyaloshinskii-Moriya (DM) interaction in the system via antisymmteric coupling. In lower symmetric perovskites such as BFO, where the $Fe^{3+}$ spin moments are already coupled through super-exchange interaction, can have the possible intrinsic existence of DM interaction.[56] According to recent first principle calculations, the evidence for DM-interaction in BFO is well proved.[57] In general, DM interaction can be expressed by a Hamiltonian $H = -D_{ij} \bullet (S_i \times S_j)$, where $D_{ij}$ is a vector which lies along a high symmetry axis and tends to couple the two spins perpendicularly.[56, 57] Due to this higher order effect the spins can be canted away from the antiferromagnetic axis by about $1°$.[56] In A-site cation disordered RE-BFO ceramic, a significant octahedral tilt coupled via octahedral rotation could give rise to significant DM interaction due to antisymmetric exchange of the form $D_{ij} \bullet S_i \times S_j$ superimposed on the spiral spin antiferromagnetic configuration stabilized by symmetric exchange of the form $J_{ij} S_i \bullet S_j$ to cant the spins noticeably.[53] This spin canting produces a competing interaction between antiferromagnetic and ferromagnetic exchange interactions which presumably also lead to origin of WFM in RE-BFO samples. Recent, experimental result based on Mossbauer spectroscopy in 15% Eu substituted BFO ceramic also supports the pertinent role of DM interaction in RE-BFO samples where the spin canting of about $0.6°$ was observed.[58] Thus the observed Weak ferromagnetism in RE-BFO samples can have the contribution due to both the mechanisms.

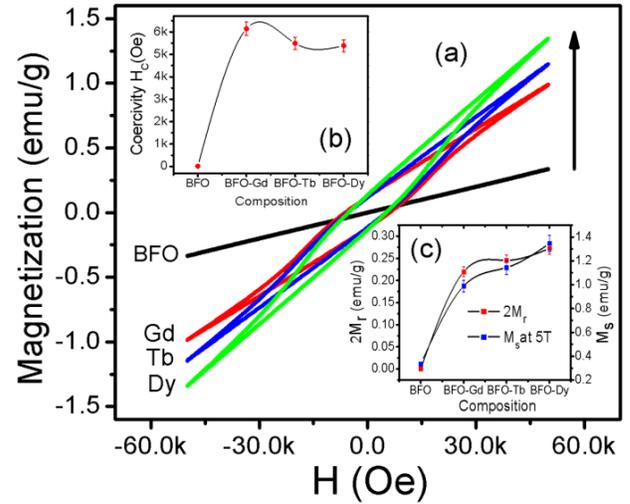

Figure 8(a) Magnetization (M-H) curves for (a) BFO, (b) BFO-Gd, (c) BFO-Tb and (d) BFO-Dy ceramics measured at 300K. Inset of Figure 8 (b) & (c) depicts the variation in coercivity ($H_C$), remanent magnetization (2$M_r$) and spontaneous magnetization ($M_S$) values at 5 Tesla with different composition.

Differential scanning calorimetric (DSC) results measured from room temperature to 450°C shown in Figure 9(a) depicts the variation in Neel transition temperature ($T_N$) with respect to RE-ion substitution in BFO lattice. The apparent λ-type transition specifies that $T_N$ in BFO and RE-BFO belongs to a second order phase transition.[54, 67] The sharp endothermic peak at ∼372.5°C in BFO sample corresponds to the Neel transition ($T_N$) temperature, which is consistent with several previous reports.[14, 22, 23, 54] RE-ion substitution with an invariable concentration of 10 mol% in BFO, shifts the $T_N$ towards the lower degree with a shift of around ∼10°C for BFO-Dy sample as shown in the inset of Figure 9(a). Further as seen in the inset of Figure 9(a), RE-BFO sample evidences a diffused peak near the Neel transition rather than the sharp peak as observed in BFO sample. This could be probably due to the coexisting orthorhombic phases in RE-BFO samples.[22] In general, $T_N$ for the antiferromagnetically ordered iron sublattice is proportional to the number of linkages (Z) per $Fe^{3+}$ ion, to the exchange constant (J<0) of $Fe^{3+}$ion pairs and to the average cosine of the angle $\theta$ between $Fe^{3+}$-$O^{2-}$-$Fe^{3+}$ linkages.[59, 60] In BFO S =5/2 for $Fe^{3+}$ ion having high spin configuration; Z=6 (i.e., due to $FeO_6$ environment) and J is negative due to super exchange interaction, hence $T_N$ can be related to the equation: $T_N \sim JZS(S+1)\cos\theta$.[59] Phenomenonlogically, according to this equation $T_N$ in BFO has a $\cos\theta$ dependence between the $Fe^{3+}$-$O^{2-}$-$Fe^{3+}$ bond angle. In RE-BFO samples due to the decrease in t value as discussed earlier, the bending angle ($\emptyset$) increases. As shown in the inset of the schematic in Figure 9(a), if the bending angle $\emptyset$ increases, $Fe^{3+}$-$O^{2-}$-$Fe^{3+}$ bond angle ($\theta$) will decrease and as result $T_N$ will decrease monotonically with $\cos\theta$.

perovskite family, the variation in phase transition $T_N$ can be also explained by following their similar empirical model. BFO has a distorted perovskite system with A-site cation ($Bi^{3+}$) of radius ($r_A^0$) and in the RE substituted BFO system, RE-ions have smaller ionic radii $r_A$ (i.e., $r_A < r_A^0$). Due to this A-site cation size variance ($\sigma_A^2$), BFO undergoes further lateral oxygen displacement i.e., $Q = r_A^0 - r_A$.[46, 62, 64]

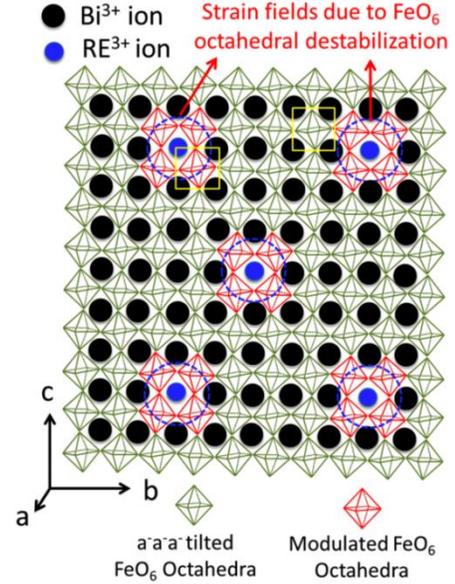

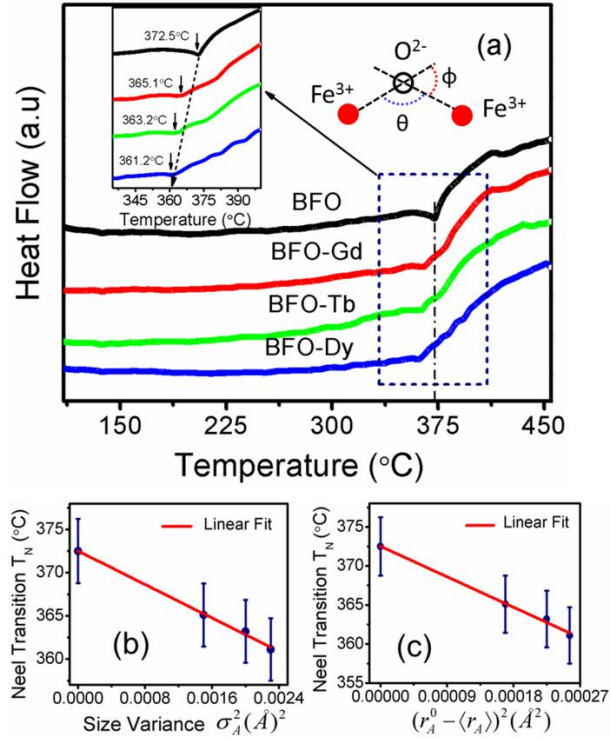

Figure 9 (a) DSC trace curves for RE-ion substituted BFO ceramic from RT-450°C, Inset of (a) shows shift in $T_N$ values with RE-ion substitution, Schematic representation for the bending angle ($\emptyset$) and $Fe^{3+}$-$O^{2-}$-$Fe^{3+}$ bond angle ($\theta$) (b) linear fit plotted between $T_N$ and Size variance ($\sigma_A^2$) (c) linear fit plotted between $T_N$ and coherent strain parameter $(r_A^0 - \langle r_A \rangle)^2$.

The change in $Fe^{3+}$-$O^{2-}$-$Fe^{3+}$ bond angle arises primarily due to the modulation in the rigid $FeO_6$ octahedra.[53] In RE-BFO samples, there is a destabilization in the anti-phase tilted ($a^-a^-a^-$) $FeO_6$ octahedra due to the oxygen displacements as evidenced from our Raman spectroscopy results. As a result, a strain field will be created around the modulated $FeO_6$ octahedral region as depicted in the schematic Figure 10. In order to provide explanation for the decrease in $T_N$ value for RE-BFO samples, we have adapted the model based on structural considerations proposed by J. P. Attfield and his co-workers[61-64] in several perovskite systems. They attributed the changes in phase transitions ($T_X$) to the strains resulting from oxygen displacements.[46, 62, 64] Since BFO also belongs to

Figure 10. Schematic representation illustrating the strain fields developed due to the ordered $FeO_6$ octahedral displacements in RE-BFO samples, along $(100)_P$ plane.

Assuming that electronic transition in RE-BFO samples i.e., $T_N$ is reduced from an ideal value of $T_N^0$ (372.5°C) by strain like interactions from oxygen displacements Q, then the transition energy $k_B T_N$ ($k_B$ is the Boltzmann constant) becomes $k_B T_N = k_B T_N^0 - \langle CQ^2 \rangle$, Where C contains the average force constant for the strain term ($\langle \rangle$ denotes an average). $\langle Q^2 \rangle$ is split into incoherent and coherent strain terms by $\langle Q^2 \rangle = \langle (r_A^0 - r_A)^2 \rangle = \sigma_A^2 + (r_A^0 - \langle r_A \rangle)^2$ and $k_B T_N$ becomes,

$k_B T_N = k_B T_N^0 - C(\sigma_A^2 + (r_A^0 - \langle r_A \rangle)^2)$. From this equation it is evident that $T_N$ will decrease linearly from its ideal value $T_N^0$ with respect to $\sigma_A^2$ and $(r_A^0 - \langle r_A \rangle)^2$. By plotting $T_N$ values against $(\sigma_A^2)$ as shown in Figure. 9(b), it depicts a linear variation with decrease in $T_N$ values. In general, slope of the linear fit $dT_x/d\sigma_A^2$ is negative for electronic transition and positive for structural phase transition and is of magnitude $\sim 10^3$-$10^4$ KÅ$^{-2}$.[46, 62, 64] In our fitting $dT_N/d\sigma_A^2$ was found to be $\sim 4848.24$ KÅ$^{-2}$ which clearly indicates that phase transition ($T_N$) strongly belongs to a second order and it is purely electronic in character (i.e., antiferro to paramagnetic).

The plot between $T_N$ against coherent strain parameter $(r_A^0 - \langle r_A \rangle)^2$ as shown in Figure 9c also gives a similar distribution. The similar distribution of $T_N$ for both coherent and in-coherent strain terms indicates that RE-ions ($Gd^{3+}$, $Tb^{3+}$, $Dy^{3+}$) substitution in BFO samples are microscopically homogeneous and as a result there is an ordered oxygen displacements[46, 64] in RE-BFO samples, which gives rise to decrease in Neel transition ($T_N$) temperatures.

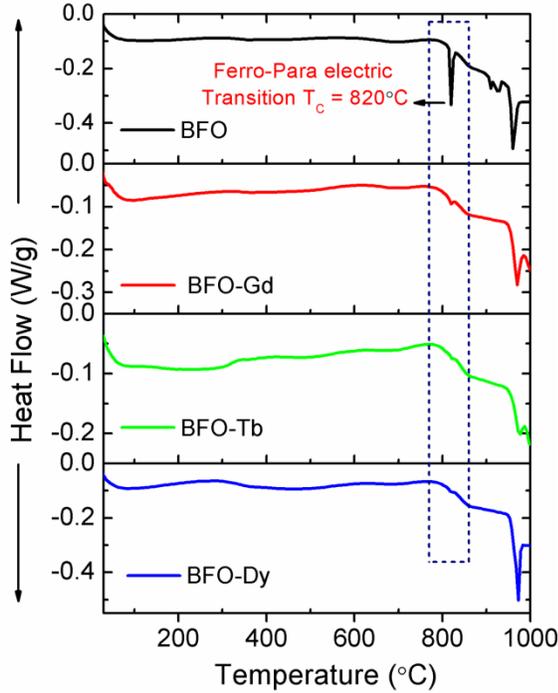

Figure 11 High temperature DSC scan showing various phase transitions of BFO and RE–BFO samples.

DSC signal recorded during heating cycle shown in Figure (11) depicts the evolution of different phase transitions in BFO and RE-BFO samples as a function of temperature upto 1000°C. From the DSC curves of BFO as shown in Figure (11) two well defined strong endothermic peaks can be seen around 820°C, 960°C and two weak endotherms around 910°C and 930°C. The observed endotherms match closely with that of the previous reports.[54, 65-67]

(i) The first strong endotherm at ∼820°C corresponds to first order ferroelectric to paraelectric transition i.e., from α-phase of $R3c$ symmetry to an orthorhombic β-phase.[65-67] The second strong endotherm at ∼959°C corresponds to the peritectic decomposition of $Bi_2Fe_4O_9$ into flux and $Fe_2O_3$.

(ii) The weak endotherm around ∼910°C corresponds to ferroelastic transitions (β to γ phase) and the second diffused endotherm at ∼930°C corresponds to peritectic decomposition of the γ-phase into flux and $Bi_2Fe_4O_9$.[65-67]

On close observation as shown in Figure 12 (a) & (b), suppression in the endotherm corresponding to Curie transition ($T_C$) and a higher degree shift in the decomposition temperature ($T_D$) were witnessed in the RE-BFO samples. The suppression of endotherm around $T_C$ in RE-BFO samples clearly implies that RE-ion substitution in BFO lattice suppresses the macroscopic ferroelectricity. The $6s^2$ lone pair mechanism in $Bi^{3+}$ ions was responsible for the origin of ferroelectricity in BFO.[3,4] The substitution of RE-ions ($Gd^{3+}$, $Tb^{3+}$, $Dy^{3+}$) has forbidden $6s^2$ lone pair which directly affects the Bi–O covalent bond, thus the hybridization. This in turn could affect the $FeO_6$ octahedral tilting due to destabilization in the ziz-zag oxygen chain.[34] The observation of significant change in the phonon mode corresponding to Bi-O bond and Oxygen displacements (i.e., $FeO_6$ modulation) evidenced through Raman spectroscopy favors this behavior.

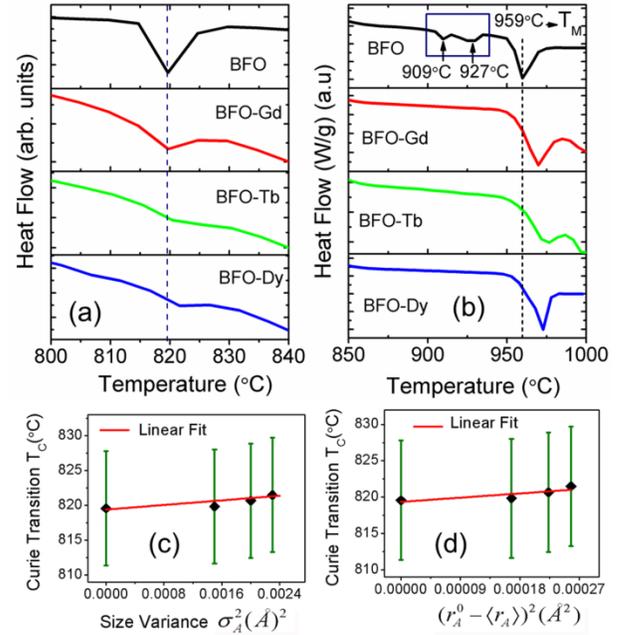

Figure 12 (a) Suppression and slight shift in the ferroelectric curie transition temperature ($T_C$) values, (b) Shift in the decomposition temperature ($T_D$) with RE-ion substitution, (c) Linear fit plotted between $T_C$ and A-site size variance and (d) Linear fit plotted between $T_C$ and coherent strain parameter and.

As a result it leads to the suppression of macroscopic ferroelectricty in RE-BFO samples. Similarly, the other possible reason for the suppression of ferroelectricity in RE-BFO samples may also be due to the orthorhombic phase coexistence along with $R3c$ phase as evidenced from XRD investigations, presumably which may belong to a non-centro symmetric space group (i.e., $Pnma$ /$Pbnm$ etc.,). Upon further close observation as depicted in Figure 12(a), an insignificant increase in $T_C$ values was also observed in RE-BFO samples. The plots between $\sigma_A^2$ vs. $T_C$ and $(r_A^0 - \langle r_A \rangle)^2$ vs. $T_C$ as shown in Figure 12 (c) & (d) exhibit a positive slope, which indicates that the transition belongs to first order type i.e., structural transition from rhombohedral to orthorhombic. Thus the explanation based on J. P. Attfields model exploits well for the changes in phase transition for the RE-BFO samples.

The frequency dependence of the real part (ε′), loss tangent (inset) and imaginary part (ε″) of dielectric constant at room temperature are plotted in Figure 13 (a) & (b). Relative permittivity ε′ of BFO is 115, while for BFO-Gd, BFO-Tb and BFO-Dy samples exhibit ε′ values of about 86, 80 and 71 respectively, at 1 kHz. It is observed that both ε′ and ε″ decrease monotonically with the frequency and attains constant value at high frequency region in all the samples. It can be explained by the phenomenon of dipole relaxation.

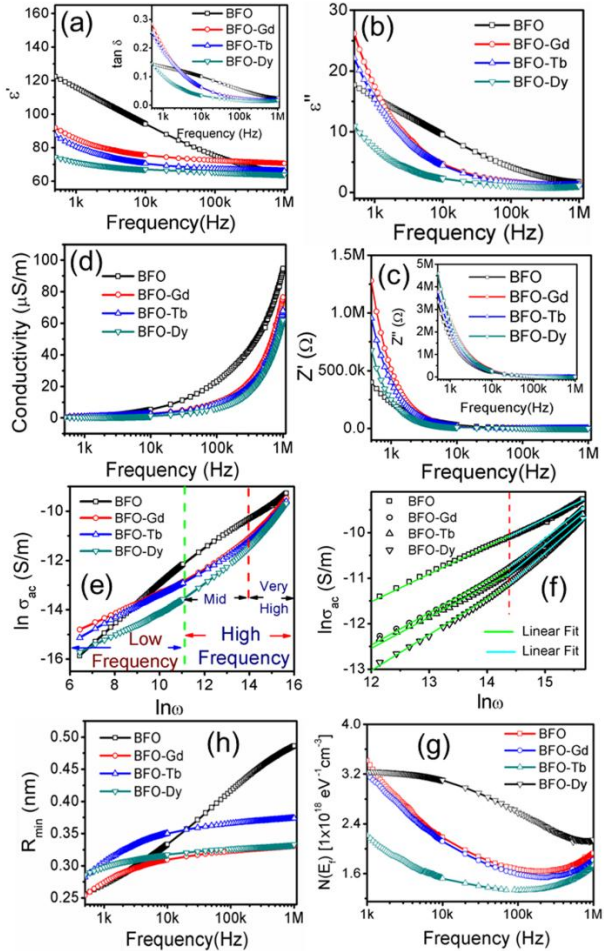

Figure. 13(a) Frequency dependence of (a) real part of permitivity (ε′), loss tangent (inset), (b) imaginary part of permitivity (ε″), (c) complex impedance Z′ and Z″ (inset) and (d) a.c. conductivity ($\sigma_{ac}$) for BFO and RE–BFO samples measured at room temperature. Figure. 13(e) Plots showing the variation between $\ln\sigma_{ac}$ and $\ln\omega$, (f) Linear fits plotted in the mid and high frequency region, (g) Density of states N(E$_f$) variation with frequency dependence and (h) Frequency dependence of hopping length R$_{min}$ for BFO and RE–BFO samples.

At low frequencies, dipoles follow the applied field and we have ε′ ≈ ε$_s$ (value of dielectric constant at quasi static fields). But as the frequency increases, dipoles begin to lag the field and ε′ decreases slightly. When the frequency reaches the characteristic frequency of a material, dielectric constant drops and at very high frequencies, dipoles can no longer follow the field and dielectric constant ε′ ≈ ε$_∞$. It is evident from Figure 13(a) that, BFO exhibits a broad dispersion in both permittivity and loss tangent which implies the dominance of space charge polarization. This may be due to possible oxygen vacancies in the BFO sample. However, RE-ion substituted BFO ceramic shows less dispersion in both permittivity and loss tangent which may be due to the possible reduction in oxygen vacancies.[68, 69] Decrease in dielectric constant with the inclusion of RE–ions ($Gd^{3+}$, $Tb^{3+}$, $Dy^{3+}$) at Bi-site in BFO sample may be due to the distortion in the crystal structure and reduction in Bi-O bond lengths at Bi-site which in turn decreases the atomic polarizability and hence the dielectric constant.[70]

Figure 13(c) shows the frequency variation of real and imaginary parts of impedance (Z′ and Z″) respectively. Decrease in Z′ with frequency indicates the presence of dielectric relaxations in these samples.[71] At low frequency all the samples depicts frequency dispersion, whereas at high frequency the value of Z′ attains a constant value. However with RE-ion substitution, increase in Z′ value is observed at low frequency region. This may be due to the decrease of grain size with the substitution of RE-ions in BFO, as it is evident from FE-SEM micrographs. Generally, in polycrystalline ceramics grains act as a semiconductor whereas grain boundaries act as insulators.[72] Due to the reduction in grain size in RE-BFO samples, grain boundary contribution dominates which gives rise to the increased Z′ values at low frequencies. Figure 13(d) shows the frequency variation of ac conductivities for BFO and RE-BFO samples. The a.c. conductivity ($\sigma_{a.c}$) of all the samples was calculated using the equation, $\sigma_{a.c} = \varepsilon_o \varepsilon \omega \tan\delta$, where ε$_o$ and ω is the permittivity of free space and angular frequency respectively. Figure 13(d) indicates that conductivity is independent of frequency up to 10 kHz for BFO sample, where as it is independent of frequency up to 100 kHz for the RE-substituted BFO samples.

According to Jonscher,[72] the frequency dependent electrical conductivity can be expressed as $\sigma_{a.c} = \sigma_{d.c} + \sigma(\omega)$, Where $\sigma_{d.c}$ represents d.c conductivity and $\sigma(\omega)$ denotes the a.c conductivity which can be represented as a power law of frequency $\sigma(\omega) = A\omega^s$, where ω is angular frequency of applied a.c field at which a.c conductivity σ is measured and s is a power law index (0 ≤ s ≤1) and A is a constant.[73] The value of s is calculated with the formula $s = d\log\sigma_{ac}/d\log\omega$.[73-76] This exponent value s is used to characterize the electrical conduction mechanism in different materials. Figure 13 (e) shows the plot between $\ln\sigma_{ac}$ and $\ln\omega$ for the entire frequency region upto 1MHz. Since there will be a contribution to the a.c. conductivity from the space charge at low frequencies below 1kHz, we have calculated the values of s from the above equation in the frequency range from 1kHz to 500kHz as shown in Figure 13(f). The value of power law exponent (s) for BFO and RE-substituted BFO samples were found to be 0.37, 0.49, 0.47 and 0.54 respectively. From the a.c conductivity plot measured at room temperature as shown in Figure 13(d), it reveals an increasing trend of $\sigma_{a.c}$ with $\omega$. This implies that $A\omega^s$ follows the mechanism based on Correlated Barrier Hopping (CBH) model.[73-76] Based on CBH model, the a.c. conductivity data was used to evaluate the density of states at Fermi level $N(E_f)$ using the following relation[73-76]

$$\sigma_{ac}(\omega) = \frac{\pi}{3} e^2 \omega k_B T [N(E_f)]^2 \alpha^{-5} [\ln(\frac{fo}{\omega})]^4$$

where e is the electronic charge, $fo$ is the photon frequency and $\alpha$ is the localized wave function. Assuming $fo = 10^{13}$ Hz, and $\alpha = 10^{10}$ m$^{-1}$ at various operating frequencies and temperatures. Figure 13(g) shows the frequency dependence of $N(E_f)$ measured at 300K. It is observed from the $N(E_f)$ plots, that RE-BFO samples may possibly have less oxygen vacancies than BFO, which is evident from the lower $N(E_f)$ values for RE-BFO samples. Further, $N(E_f)$ values tend to decrease for all samples with the frequency and increases slightly at higher frequency regime, because $N(E_f)$ is dependent on $\omega$ both in the upper and lower terms $N(E_f) = \sqrt{(\ln \omega)^4 / \omega}$. Hopping conduction mechanism occurs generally, in the materials having a band gap similar to that of a semiconductor along with the existence of a high density of states $N(E_f)$ of charge carriers.[75] Since BFO has a band gap value close to that of the semiconductor range and lies between ~2.5eV-2.8eV,[48-51] with high density of charge carriers, hence the conduction in BFO can be considered due to the hopping of charge carriers between the nearest neighbors.[75] To support this behavior, the minimum hopping length $R_{min}$ was estimated using the relation $R_{min} = 2e^2 / \pi \varepsilon \varepsilon_o W_m$, where $W_m = 6k_B T / (1-s)$ is the binding energy.[75,76] Figure 13(h) shows the variation of $R_{min}$ (in the order about ~$10^{-10}$m) with frequency at room temperature for all samples. The $R_{min}$ plot elucidates that BFO shows a low $R_{min}$ value in the low frequency region and it displays a continuous dispersion with increase in $R_{min}$ values upto high frequency region. However, it tries to saturate in the high frequency region close to 1MHz (still high frequency is required to see the exact saturation), but in the RE-BFO samples $R_{min}$ value tends to saturate even in the mid frequency region around 10kHz. This behavior supports the long range mobility of charge carriers in RE-BFO samples. On the other hand, the sigmoidal increase in the value of $R_{min}$ in BFO sample with frequency can be attributed to the conduction phenomena due to short-range mobility of charge carriers. This supports well with the high density of charge carriers for BFO as observed from the $N(E_f)$ plots in Figure 13(h). As a summary from the above $N(E_f)$ and $R_{min}$ results we can attribute that RE-BFO samples exhibits less number of vacant sites than the BFO sample.

## CONCLUSION

In summary, our present experimental study demonstrates that A-site cationic size variance and disorder effects dramatically influence the physical properties of $Bi_{0.9}RE_{0.1}FeO_3$ ceramics. As a result of large cationic size variance and lower tolerance factor values in RE-BFO sample favors the formation of coexisting orthorhombic phases along with $R3c$. RE-ion substitution in BFO induces a change in the $FeO_6$ local environment in terms of their crystal filed splitting due to the increase in internal chemical pressure as revealed from spectroscopic studies. The RE-ion substitution in BFO lattice disturbs the covalent character between Bi-O bond and also considerably affects the stability of the $FeO_6$ octahedral chains, which in turn suppresses the macroscopic ferroelectricity. A weak ferromagnetism (WFM) observed in this study strongly confirms the existence of D-M interaction due to competing antiferromagnetic and ferromagnetic interactions in RE-BFO system. The ordered oxygen displacements due to RE-ion substitution give rise to strain fields which significantly lowers the Neel transition temperature $T_N$. The nature of the phase transitions has been explained on the basis of coherent and incoherent strains developed within the lattice. RE-ion substitution in BFO lattice also reduces the vacant sites as observed from the hopping length and Density of states values.

## ACKNOWLEGEMENT

Authors are grateful to the Department of Science and Technology (DST), Government of India for their financial support under Fast Track scheme (SR/FTP/PS-065/2011) to carry out this work. The authors also thank Dr. Buchi Suresh, ARCI Hyderabad, for availing their high temperature DSC measurements, Dr. R. Ranjith, IIT Hyderabad for his valuable technical discussions. Authors gratefully acknowledge Defence Metallurgical Research Laboratory (DMRL) and Indian Institute of Technology Hyderabad (IITH), India for their extended support to carry out this work.